\documentclass[preprints,article,accept,moreauthors,pdftex]{mdpi} 

\firstpage{1} 
\pubvolume{xx}
\issuenum{1}
\articlenumber{5}
\pubyear{2020}
\copyrightyear{2020}
\history{}
\updates{} 

\Title{Precise Half-Life Values for Two-Neutrino Double-$\beta$ Decay: 2020 review}

\Author{{Alexander Barabash} *}

\AuthorNames{Alexander Barabash}

\address[1]{National Research Centre ``Kurchatov Institute'', Institute of Theoretical and Experimental Physics, B.~Cheremushkinskaya 25, 117218~ Moscow, Russia; barabash@itep.ru; Tel.: +7-916-933-9350}

\abstract{All existing positive results on two-neutrino double beta decay and two-neutrino double electron capture in different nuclei have been analyzed. Weighted average and recommended  half-life values for 
$^{48}$Ca, $^{76}$Ge, $^{82}$Se, $^{96}$Zr, $^{100}$Mo, $^{100}$Mo - 
$^{100}$Ru ($0^+_1$), $^{116}$Cd, $^{128}$Te, $^{130}$Te, $^{136}$Xe, $^{150}$Nd, $^{150}$Nd - $^{150}$Sm 
($0^+_1$), $^{238}$U, $^{78}$Kr, $^{124}$Xe and $^{130}$Ba have been obtained. Given the measured half-life values, effective nuclear matrix elements for all these transitions 
were calculated.}


\begin{document}


\section{Introduction}
\label{sec.1}
Two-neutrino double beta decay ($2\nu\beta\beta$) was first considered by Maria Goeppert-Mayer in 1935~\cite{GOE35}:

\begin{equation}
(A,Z) \rightarrow (A,Z+2) + 2e^{-} + 2\bar { \nu}
\end{equation}                                 

This is a process in which a nucleus (A,Z) decays to a nucleus (A,Z + 2) by emitting two electrons and two electron-type antineutrinos. The $2\nu\beta\beta$ decay is a second-order weak interaction process and does not violate any conservation laws. Nevertheless, the study of this process provides rich information that can be used both to clarify various aspects of neutrinoless double beta decay and to search for exotic processes (decays with Majoron emission \cite{ARN06,ARN19}, bosonic neutrinos \cite{BAR07}, violation of Lorentz invariance \cite{ARN19,ALB16,AZO19a}, the presence of right-handed leptonic currents \cite{DEP20}, neutrino self-interactions ($\nu$SI) \cite{DEP20a},  etc.). The $2\nu\beta\beta$ decay was first discovered in a geochemical experiment with $^{130}$Te in 1950~\cite{ING50}. In a direct (counter) experiment, the decay was first recorded by M. Moe et al. in 1987 (TRC, 
 $^{82}$Se)~\cite{ELL87}. To date, $2\nu\beta\beta$ decay has already been studied quite well. This process has been registered for 11 nuclei. For some nuclei ($^{100}$Mo, $^{150}$Nd), a transition to the 0$^+_1$ excited state of the daughter nucleus was detected too. In addition, a two-neutrino double electron capture (ECEC(2$\nu$)) was detected in several nuclei ($^{130}$Ba \cite{MES01}, $^{124}$Xe \cite{APR19}, $^{78}$Kr \cite{RAT17}). In this process, two orbital electrons are captured. In the final state, two neutrinos and two X-rays  appear:

\begin{equation}
e^{-} + e^{-} + (A,Z) \rightarrow (A,Z-2) + 2\nu + 2X.
\end{equation}    

In the NEMO-3 
experiment, all decay characteristics (total energy spectrum, single electron spectrum, angular distribution) for 7 isotopes ($^{48}$Ca, $^{82}$Se, $^{96}$Zr, $^{100}$Mo, $^{116}$Cd, $^{130}$Te and $^{150}$Nd) were studied simultaneously. At present, the study of two-neutrino processes is moving into a new stage, precision study. The accuracy of determining the half-life values and other characteristics of this process is becoming increasingly important (see the discussion in \cite{ARN19,DEP20,BAR10,BAR15,SIM18}). The exact half-lives are important to know for the following reasons:

1. {\it Nuclear spectroscopy}. 
It has now been established that some isotopes that were previously considered stable are not, and decay of these isotopes is observed through the $2\nu\beta\beta$ decay with a half-life of
$\sim 10^{18} - 10^{24}$ yr. One just need to know the exact half-life values to include them in the isotope tables.  Then, these values can be used for any purpose.

2. {\it Nuclear matrix elements (NME)}. First of all, one can check the quality of
NME calculations for $2\nu\beta\beta$ decay, because it is possible to directly compare experimental and calculated values. Secondly, accurate knowledge of the NME(2$\nu$) also makes it possible to improve the quality of NME calculations for neutrinoless double beta decay ($0\nu\beta\beta$). For example, the accurate half-life values for $2\nu\beta\beta$ decay are used to determine the most important parameter of the quasiparticle random-phase approximation model (QRPA), the strength of the particle–particle interaction $g_{pp}$~ \cite{ROD06,ROD07,KOR07,SIM08}.

3. {\it To fix $g_A$ (axial-vector coupling constant)}. There are indications that, in nuclear medium, the matrix elements of the axial-vector operator are reduced in comparison with their free nucleon values. This~quenching is described as a reduction of the coupling constant $g_A$ from its free nucleon value of $g_A$ = 1.2701 \cite{BER12} to the value of $g_A$ $\sim$ 0.35-1.0 (see \cite{ENG17,SUH17,SUH19}). In principle, $g_A$ value could be established by comparison of exact experimental values and results of theoretical calculations of NMEs. Finally,~it~can help in understanding the $g_A$ value in the case of $0\nu\beta\beta$ decay (see discussions in Refs.~ \cite{ENG17,SUH17,SUH19}).

It should be noted here that the phenomenological interpretation of the change in the value of $g_A$ in nuclear matter is apparently connected with the imperfection of our description of the nuclear structure and the process of double beta decay itself. Therefore, when describing the process of $2\nu\beta\beta$ decay, we are essentially adjusting the value of $g_A$ in order to give a correct description of the process. In this sense, this is about the same as the situation with $g_{pp}$ in the previous paragraph.

4. {\it A check of the “bosonic” neutrino hypothesis \cite{BAR07} and $\nu$SI \cite{DEP20a}.}
 
At the same time, it is quite difficult to choose the "best" result from the available data. For~some isotopes, up to 7--10 different measurements exist. The quality of these results is not always obvious. Therefore, it is difficult to choose the best 
 (``correct'') value for the half-life.

In the present paper, a critical analysis of all available results on two-neutrino decay has been performed and average or/and recommended half-life values for all isotopes are presented. Using these values and the values of the phase space factors from \cite{KOT12,MIR15}, the ``effective'' NMEs were~calculated.

The first time that such type of work was done was in 2001, and the 
results were presented at the International Workshop on the calculation of double beta decay 
nuclear matrix elements, MEDEX'01~\cite{BAR02}. Then, updated half-life values were
presented at MEDEX'05, MEDEX'09 and MEDEX'13 and published in 
Refs. \cite{BAR06}, \cite{BAR10} and \cite{BAR15}, respectively.  
In this article, new positive results obtained since the beginning of 2015 
and to the middle of 2020 have been added and analyzed. Preliminary~results of this analysis have been presented at MEDEX'19 \cite{BAR19}.

The main differences from the previous analysis \cite{BAR15} are the following:
(1) The new experimental results are included in the analysis: $^{48}$Ca \cite{ARN16}, 
$^{76}$Ge \cite{AGO15}, $^{82}$Se \cite{ARN18,AZZ19}, $^{100}$Mo \cite{ARN19,ARM20}, 
$^{116}$Cd \cite{ARN17,BAR18}, 
$^{130}$Te~\cite{ALD17,NUT20}, $^{136}$Xe \cite{GAN16}, $^{150}$Nd \cite{ARN16a}, $^{150}$Nd - $^{150}$Sm (0$^+_1$) \cite{KAS19} and $^{130}$Ba~\cite{MES17};
(2) the positive results obtained for $^{78}$Kr \cite{RAT17} and $^{124}$Xe \cite{APR19} are added (these decays have been detected for the first time). I would like to stress that most of the above-mentioned new results are very precise. The accuracy of some of the obtained half-life values is $\sim$2--3\%. This is a result of (mainly) experiments with low-temperature bolometers (including scintillating bolometers) 
 
\section{Experimental Data}

Tables \ref{Table1} and \ref{Table2} show the experimental results on $2\nu\beta\beta$ decay and on ECEC(2$\nu$) capture in different nuclei. For direct experiments, the number of detected (useful) events and the signal-to-background (S/B) ratio are presented.

\begin{table}[H]
\caption{Present, positive $2\nu\beta\beta$ decay results. 
N is the number of useful events, S/B is the signal-to-background 
ratio.}
\centering
\label{Table1}
\begin{tabular}{ccccc}
\toprule
\textbf{Nucleus} & \textbf{N} & \textbf{$T_{1/2}$, yr} & \textbf{S/B} & \textbf{Ref., Year} \\
\midrule

$^{48}$Ca & $\sim100$ & $[4.3^{+2.4}_{-1.1}(stat)\pm 1.4(syst)]\cdot 10^{19}$
  & 1/5 & \cite{BAL96}, 1996 \\
 & 5 & $4.2^{+3.3}_{-1.3}\cdot 10^{19}$ & 5/0 & \cite{BRU00}, 2000 \\
& 116 & $[6.4^{+0.7}_{-0.6}(stat)^{+1.2}_{-0.9}(syst)]\cdot 10^{19}$ & 3.9 & \cite{ARN16}, 2016 \\
\midrule
 & & {\bf Average value:} $\bf 5.3^{+1.2}_{-0.8} \cdot 10^{19}$ & & \\  
\midrule
$^{76}$Ge & $\sim4000$ & $(0.9\pm 0.1)\cdot 10^{21}$ & $\sim$                                                  1/8
& \cite{VAS90}, 1990 \\
& 758 & $1.1^{+0.6}_{-0.3}\cdot 10^{21}$ & $\sim 1/6$ & \cite{MIL91}, 1991 \\
& $\sim330$ & $0.92^{+0.07}_{-0.04}\cdot 10^{21}$ & $\sim$1.2 & \cite{AVI91}, 1991 \\
& 132 & $1.27^{+0.21}_{-0.16}\cdot 10^{21}$ & $\sim 1.4$ & \cite{AVI94}, 1994 \\
& $\sim3000$ & $(1.45\pm 0.15)\cdot 10^{21}$ & $\sim$1.5 & \cite{MOR99}, 1999 
\\
& $\sim80,000$ & $[1.74\pm 0.01(stat)^{+0.18}_{-0.16}(syst)]\cdot 10^{21}$ & $\sim$1.5 
& \cite{HM03}, 2003 \\
& 25,690 & $(1.925\pm 0.094)\cdot 10^{21}$ & $\sim$3 & \cite{AGO15}, 2015 \\
\midrule
& & {\bf Average value:} $\bf (1.88\pm 0.08)\cdot 10^{21}$ & & \\  
\midrule

$^{82}$Se& 89.6 & $1.08^{+0.26}_{-0.06}\cdot 10^{20}$ & $\sim 8$ & \cite{ELL92}, 1992 \\
 & 149.1 & $[0.83 \pm 0.10(stat) \pm 0.07(syst)]\cdot 10^{20}$ & 2.3 & 
\cite{ARN98}, 1998 \\
& 2750 & $[0.939 \pm 0.017(stat) \pm 0.058(syst)]\cdot 10^{20~(a)}$ & 4 & \cite{ARN18}, 
2018\\ 
& $\sim$200,000 & $[0.860 \pm 0.003(stat) ^{+0.019}_{-0.013}(syst)]\cdot 10^{20~(a)}$ & $\sim$10 & \cite{AZZ19}, 
2019\\ 
& & $(1.3\pm 0.05)\cdot 10^{20}$ (geochem.) & & \cite{KIR86}, 1986 \\
\midrule
& & {\bf Average value:} $\bf 0.87^{+0.02}_{-0.01}\cdot 10^{20}$ & & \\
\midrule
$^{96}$Zr & 26.7 & $[2.1^{+0.8}_{-0.4}(stat) \pm 0.2(syst)]\cdot 10^{19}$ & $1.9^{~(b)}$ 
& \cite{ARN99}, 1999 \\
& 453 & $[2.35 \pm 0.14(stat) \pm 0.16(syst)]\cdot 10^{19}$ & 1 & \cite{ARG10}, 2010\\
& & $(3.9\pm 0.9)\cdot 10^{19}$ (geochem.)& & \cite{KAW93}, 1993 \\
& & $(0.94\pm 0.32)\cdot 10^{19}$ (geochem.)& & \cite{WIE01}, 2001 \\
\midrule
& & {\bf Average value:} $\bf (2.3 \pm 0.2)\cdot 10^{19}$ & & \\ 
\midrule
$^{100}$Mo & $\sim$500 & $11.5^{+3.0}_{-2.0}\cdot 10^{18}$ & 1/7 & 
\cite{EJI91}, 1991 \\
& 67 & $11.6^{+3.4}_{-0.8}\cdot 10^{18}$ & 7 & \cite{ELL91}, 1991 \\
& 1433 & $[7.3 \pm 0.35(stat) \pm 0.8(syst)]\cdot 10^{18~(a)(c)}$ & 3 & 
\cite{DAS95}, 1995 \\
& 175 & $7.6^{+2.2}_{-1.4}\cdot 10^{18}$ & 1/2 & \cite{ALS97}, 1997 \\
& 377 & $[6.82^{+0.38}_{-0.53}(stat) \pm 0.68(syst)]\cdot 10^{18}$ & 10 & 
\cite{DES97}, 1997 \\
& 800 & $[7.2 \pm 1.1(stat) \pm 1.8(syst)]\cdot 10^{18}$ & 1/9 & 
\cite{ASH01}, 2001 \\
& $\sim$350 & $[7.15 \pm 0.37(stat) \pm 0.66(syst)]\cdot 10^{18}$ & $\sim$ $5^{~(d)}$ & 
\cite{CAR14}, 2014\\
& 500,000 & $[6.81 \pm 0.01(stat)^{+0.38}_{-0.40}(syst)]\cdot 10^{18~(a)}$ & 80 & 
\cite{ARN19}, 2019\\
& 35,638 & $[7.12^{+0.18}_{-0.14}(stat) \pm 0.10(syst)]\cdot 10^{18~(a)}$ & 10 & 
\cite{ARM20}, 2020\\
& & $(2.1\pm 0.3)\cdot 10^{18}$ (geochem.)& & \cite{HID04}, 2004 \\ 
\midrule
& & {\bf Average value:} $\bf 7.06^{+0.15}_{-0.13}\cdot 10^{18}$ & & \\
\bottomrule
\end{tabular}
\end{table}

\addtocounter{table}{-1}
\begin{table}
\caption{{\em Cont.}} 
\centering
\label{Table1}
\begin{tabular}{ccccc}
\toprule
\textbf{Nucleus} & \textbf{N} & \textbf{$T_{1/2}$, yr} & \textbf{S/B} & \textbf{Ref., Year} \\
\midrule
$^{100}$Mo - & $133^{~(e)}$ & $6.1^{+1.8}_{-1.1}\cdot 10^{20}$ & 1/7 & 
\cite{BAR95}, 1995 \\
$^{100}$Ru ($0^+_1$) &  $153^{~(e)}$ & $[9.3^{+2.8}_{-1.7}(stat) \pm 1.4(syst)]\cdot 
10^{20}$ & 1/4 & \cite{BAR99}, 1999 \\
 & 19.5 & $[5.9^{+1.7}_{-1.1}(stat) \pm 0.6(syst)]\cdot 10^{20}$ & $\sim$8 & 
\cite{DEB01}, 2001 \\ 
& 35.5 & $[5.5^{+1.2}_{-0.8}(stat) \pm 0.3(syst)]\cdot 10^{20}$ & $\sim$8 & 
\cite{KID09}, 2009 \\ 
& 37.5 & $[5.7^{+1.3}_{-0.9}(stat) \pm 0.8(syst)]\cdot 10^{20}$ & $\sim$3 & 
\cite{ARN07}, 2007 \\ 
& $597^{~(e)}$ & $[6.9^{+1.0}_{-0.8}(stat) \pm 0.7(syst)]\cdot 10^{20}$ & $\sim$1/10 & 
\cite{BEL10}, 2010 \\
& $239^{~(e)}$ & $[7.5 \pm 0.6(stat) \pm 0.6(syst)]\cdot 10^{20}$ & 2 & 
\cite{ARN14}, 2014 \\     
\midrule
& & {\bf Average value:} $\bf 6.7^{+0.5}_{-0.4}\cdot 10^{20}$ & & \\
\midrule
$^{116}$Cd& $\sim180$ & $2.6^{+0.9}_{-0.5}\cdot 10^{19}$ & $\sim$1/4 & 
\cite{EJI95}, 1995 \\
& 174.6 & $[2.9 \pm 0.3(stat) \pm 0.2(syst)]\cdot 10^{19~(a)(c)}$ & 3 & 
\cite{ARN96}, 1996 \\
& 9850 & $[2.9\pm 0.06(stat)^{+0.4}_{-0.3}(syst)]\cdot 10^{19}$ & $\sim$3 & 
\cite{DAN03}, 2003 \\
& 4968 & $[2.74 \pm 0.04(stat) \pm 0.18(syst)]\cdot 10^{19~(a)}$ & 12 & \cite{ARN17}, 2017\\
& 93,000 & $2.63^{+0.11}_{-0.12}\cdot 10^{19}$ & 1.5 & \cite{BAR18}, 2018\\
\midrule
& & {\bf Average value:} $\bf (2.69 \pm 0.09)\cdot 10^{19}$ & & \\
\midrule
$^{128}$Te& & $\sim 2.2\cdot 10^{24}$ (geochem.) & & \cite{MAN91}, 1991 \\
& & $(7.7\pm 0.4)\cdot 10^{24}$ (geochem.)& & \cite{BER93}, 1993 \\
& & $(2.41\pm 0.39)\cdot 10^{24}$ (geochem.)& & \cite{MES08}, 2008 \\
& & $(2.3\pm 0.3)\cdot 10^{24}$ (geochem.)& & \cite{THO08}, 2008 \\
\midrule
& & {\bf Recommended value: $\bf (2.25\pm 0.09)\cdot 10^{24~(f)}$}  & & \\
\midrule
$^{130}$Te& 260 & $[6.1 \pm 1.4(stat)^{+2.9}_{-3.5}(syst)]\cdot 10^{20}$ & 1/8 & \cite{ARN03}, 2003 \\
& 236 & $[7.0 \pm 0.9(stat) \pm 1.1(syst)]\cdot 10^{20}$ & 1/3 & \cite{ARN11}, 2011 \\
& $\sim$33,000 & $[8.2 \pm 0.2(stat) \pm 0.6(syst)]\cdot 10^{20}$ & 0.1--0.5 & \cite{ALD17}, 2017 \\
& $\sim$20,000 & $[7.9 \pm 0.1(stat) \pm 0.2(syst)]\cdot 10^{20}$ & $>$1 & \cite{NUT20}, 2020 \\
& & $\sim 8\cdot 10^{20}$ (geochem.) & & \cite{MAN91}, 1991 \\
& & $(27\pm 1)\cdot 10^{20}$ (geochem.)& & \cite{BER93}, 1993 \\
& & $(9.0\pm 1.4)\cdot 10^{20}$ (geochem.)& & \cite{MES08}, 2008 \\
& & $(8.0\pm 1.1)\cdot 10^{20}$ (geochem.)& & \cite{THO08}, 2008 \\
\midrule
& & {\bf Average value:} $\bf (7.91 \pm 0.21)\cdot 10^{20}$ & & \\
\midrule
$^{136}$Xe & $\sim$19,000 & $[2.165 \pm 0.016(stat) \pm 0.059(syst)]\cdot 10^{21}$ & 
$\sim$10 & \cite{ALB14}, 2014 \\
& $\sim$100,000 & $[2.21 \pm 0.02(stat) \pm 0.07(syst)]\cdot 10^{21}$ & $\sim$10 & 
\cite{GAN16}, 2016 \\

\midrule
& & {\bf Average value:} $\bf(2.18\pm 0.05)\cdot 10^{21}$ & & \\
\midrule
$^{150}$Nd& 23 & $[18.8^{+6.9}_{-3.9}(stat) \pm 1.9(syst)]\cdot 10^{18}$ & 
1.8 & \cite{ART95}, 1995 \\
& 414 & $[6.75^{+0.37}_{-0.42}(stat) \pm 0.68(syst)]\cdot 10^{18}$ & 6 & 
\cite{DES97}, 1997 \\
& 2214 & $[9.34 \pm 0.22(stat) ^{+0.62}_{-0.60}(syst)]\cdot 10^{18}$ & 4 & \cite{ARN16a}, 2016\\
\midrule
& & {\bf Average value:} $\bf(8.4\pm 1.1)\cdot 10^{18}$ & & \\
\midrule
& & {\bf Recommended value: $\bf(9.34 \pm 0.65)\cdot 10^{18}$} & & \\
\midrule
$^{150}$Nd - & $177.5^{~(e)}$ & $[1.33^{+0.36}_{-0.23}(stat)^{+0.27}_{-0.13}(syst)]\cdot 10^{20}$ & 
1/5 & \cite{BAR09}, 2009 \\
 $^{150}$Sm ($0^+_1$) & 21.6 & $[1.07^{+0.45}_{-0.25}(stat) \pm {+0.07}(syst)]\cdot 10^{20}$ & $\sim$1.2 & \cite{KID14}, 2014\\
& $\sim$6 & $[0.69^{+0.40}_{-0.19}(stat) \pm {+0.11}(syst)]\cdot 10^{20}$ & $\sim$2 & \cite{KAS19}, 2019\\
\midrule
 & & {\bf Average value:} $\bf 1.2^{+0.3}_{-0.2}\cdot 10^{20}$ & \\ 
 \midrule
$^{238}$U& & $\bf (2.0 \pm 0.6)\cdot 10^{21}$ (radiochem.) &  & \cite{TUR91}, 1991 \\

\bottomrule
\end{tabular}
 \begin{tabular}{c}
\multicolumn{1}{p{\linewidth-1.5cm}}{\footnotesize \justifyorcenter{$^{(a)}$ {For SSD 
 mechanism}. $^{(b)}$ For $E_{2e} > 1.2$ MeV. $^{(c)}$ After correction (see \cite{BAR10}). $^{(d)}$ For $E_{2e} > 1.5$ MeV. 
 $^{(e)}$ In both peaks. $^{(f)}$ This value was obtained using average $T_{1/2}$ for $^{130}$Te and well-known ratio $T_{1/2}$($^{130}$Te)/$T_{1/2}$($^{128}$Te) = $(3.52\pm 0.11)\cdot 10^{-4}$  \cite{BER93}.}}
\end{tabular}

\end{table}

\begin{table}[H]
\caption{Present, positive two-neutrino double electron capture  results. 
N is the number of useful events, S/B is the signal-to-background 
ratio. In the case of $^{78}$Kr and $^{124}$Xe $T_{1/2}$ for $2K(2\nu)$, capture is presented (this is $\sim$75--80\% of $ECEC(2\nu)$). }
\centering
\label{Table2}
\begin{tabular}{ccccc}
\toprule
\textbf{Nucleus} & \textbf{N}  & \textbf{$T_{1/2}(2\nu)$, yr}   & \textbf{S/B} & \textbf{Ref., Year}  \\ 
\midrule

{$^{130}$Ba} &  & {$\bf 2.1^{+3.0}_{-0.8} \cdot 10^{21}$ (geochem.)} & 
 & {\cite{BAR96}, 1996} \\
 $ECEC(2\nu)$&  & $\bf (2.2 \pm 0.5)\cdot 10^{21}$ (geochem.) & 
 & \cite{MES01}, 2001 \\
& & $\bf (0.60 \pm 0.11)\cdot 10^{21}$ (geochem.) &
& \cite{PUJ09}, 2009 \\
\midrule
& & {\bf Recommended value:} $\bf (2.2 \pm 0.5)\cdot 10^{21}$  & & \\
\midrule
$^{78}$Kr & 15 & $\bf [1.9^{+1.3}_{-0.7}(stat) \pm 0.3(syst)] \cdot 10^{22}$ & 15 
 & \cite{RAT17}, 2017 \\
$2K(2\nu)$ & & & & \\
\midrule
& & {\bf Recommended value:} $\bf (1.9^{+1.3}_{-0.8})\cdot 10^{22}$ (?
)$^{~(a)}$ &  & \\
\midrule
$^{124}$Xe & 126 & $\bf [1.8 \pm 0.5(stat) \pm 0.1(syst)] \cdot 10^{22}$ & 0.2 
 & \cite{APR19}, 2019 \\
$2K(2\nu)$ & & & & \\
\midrule
& & {\bf Recommended value:} $\bf (1.8 \pm 0.5)\cdot 10^{22}$  & & \\
\bottomrule

\end{tabular}

 \begin{tabular}{c}
\multicolumn{1}{p{\linewidth-1.5cm}}{\footnotesize \justifyorcenter{$^{(a)}$ {See text.}}}
\end{tabular}
\end{table}

\section{Data Analysis}

To calculate an average of the ensemble of available data, a standard procedure, as recommended by the Particle Data Group 
\cite{BER12}, was used.  The weighted average and the corresponding error 
were calculated, as follows:
\begin{equation}
\bar x\pm \delta \bar x = \sum w_ix_i/\sum w_i \pm (\sum w_i)^{-1/2} , 
\end{equation} 
where $w_i = 1/(\delta x_i)^2$.  Here, $x_i$ and $\delta x_i$ are 
the value and error reported by the i-th experiment, and 
the summations run over the N experiments.  

Then, it is necessary to calculate $\chi^2 = \sum w_i(\bar x - x_i)^2$ and 
compare it with N - 1, which is the expectation value of $\chi^2$ if the 
measurements are from a Gaussian distribution. In the case when $\chi^2/(N - 1)$ is 
less than or equal to 1 and there are no known problems with the data, 
then one accepts the results. In the case when $\chi^2/(N - 1) >> 1$, 
one chooses 
not to use the average procedure at all. 
Finally, if~$\chi^2/(N - 1)$ is larger than 1, but not greatly so, it is  
still best to use the average data, but to increase the quoted error, $\delta \bar x$ 
in Equation 1, by a factor of S defined by 
\begin{equation}
S = [\chi^2/(N - 1)]^{1/2}.
\end{equation} 

For averages, the statistical and systematic errors are treated in quadrature 
and used as a  combined error $\delta x_i$. In some cases, only the results 
obtained with a high enough 
S/B ratio were used.


\subsection{$^{48}$Ca}
The $2\nu\beta\beta$ decay of $^{48}$Ca was observed in three independent experiments \cite{BAL96,BRU00,ARN16}. 
The obtained results are in good agreement. The weighted average value is:
$$
T_{1/2} = 5.3^{+1.2}_{-0.8} \cdot 10^{19} \rm{yr}.
$$ 

This value is slightly higher than the average value obtained in previous analysis
\linebreak  ($T_{1/2} = 4.4^{+0.6}_{-0.5} \cdot 10^{19}$ yr \cite{BAR15}).
This is due to the fact that the final result of the NEMO-3 experiment \cite{ARN16} was used in present analysis (the intermediate result of the NEMO-3 experiment \cite{BAR11a} was used in \cite{BAR15}). The change in the final result in the NEMO-3 experiment was mainly due to the fact that after disassembling the detector, the parameters of sources containing $^{48}$Ca were refined. It was found that, in reality, the diameter of the sources turned out to be slightly larger (and the thickness, respectively, less) than previously assumed. Taking this circumstance into account led to an increase in the calculated efficiency of recording useful events and, ultimately, to an increase in the $T_{1/2}$ value for $^{48}$Ca. In addition, systematic error in \cite{ARN16} is higher then in \cite{BAR11a}.

\subsection{$^{76}$Ge } 

For $^{76}$Ge, a lot of positive results were obtained, but the scatter of the obtained values is rather large. Half-life values gradually increased over time during the 90-th. It was decided not to use the results of the early works (1990s
), as a recent historical review \cite{AVI19} emphasized that the contribution of background processes was underestimated in these works. Therefore, to determine the average value, the results published after 2000 have been used, with large statistics and a high S/B ratio \cite{HM03, AGO15}. Note that the final result of the Heidelberg--Moscow collaboration was used \cite{HM03}. As a result, we~get:

$$
    T_{1/2} = (1.88 \pm 0.08) \cdot 10^{21} \rm{yr}.
$$

\subsection{$^{82}$Se} 
\label{sec.3.3}

There are many geochemical measurements ($\sim$20) and only  four independent counting experiments for $^{82}$Se. However, the geochemical results are in poor agreement with each other and with the results of direct experiments. It is known that the possibility of existing large systematic errors in geochemical measurements cannot be excluded (see discussion in Ref. \cite{MAN86}).  Thus,  only the results of the direct measurements \cite{ELL92,ARN98,ARN18,AZZ19} were used  to obtain a present half-life value for $^{82}$Se.
Single State Dominance (SSD) mechanism (see explanation in \cite{DOM05}) was established for $2\nu\beta\beta$ transition in $^{82}$Se \cite{ARN18,AZZ19} and half-life values in this papers were obtained under the assumption of the SSD mechanism \footnote{It was experimentally demonstrated that in some nuclei ($^{82}$Se, $^{100}$Mo and $^{116}$Cd) the SSD mechanism is realized. In this case, the spectra (total energy, single electron energy and angular distribution) differ from the case of the High State Dominance (HSD) mechanism. In principle, this does not affect the half-life of the corresponding nuclei. In a real experiment, energy is recorded with a certain threshold, which can affect the efficiency of recording useful events. The neglect of this effect can lead to an error in the determination of $T_{1/2}$ (up to $\sim$ 10-15\%). This is especially noticeable in experiments where the energy of an individual electron is recorded (for example, the NEMO-3 experiment).}. 
The result of Ref. \cite{ELL87} has not been used in the analysis because this is the preliminary result of \cite{ELL92}.
The result of work \cite{ELL92} is presented with very 
asymmetrical errors. To be more conservative, the value for the lower error was taken to be the same as the upper one in our analysis. Finally, the weighted average value is:
$$
T_{1/2} = 0.87^{+0.02}_{-0.01} \cdot 10^{20} \rm{yr}.
$$ 

\subsection{$^{96}$Zr} 
There are two positive results from the direct experiments (NEMO-2 \cite{ARN99} and 
NEMO-3 \cite{ARG10}) and two geochemical results \cite{KAW93,WIE01}.  Taking into account the comment in 
Section \ref{sec.3.3}, the values from direct experiments (Refs. \cite{ARN99,ARG10}) were used to obtain 
a present weighted half-life value for $^{96}$Zr: 
$$
T_{1/2} = (2.3 \pm 0.2)\cdot 10^{19} \rm{yr}.                    
$$ 

\subsection{$^{100}$Mo}
 
By the present nine positive results from direct experiments\footnote{I do not consider here the result 
of Ref. \cite {VAS90a} because of a high background contribution 
that was 
not excluded in this experiment. As a result, the ``positive'' effect is mainly associated with the background. Calculations show that without the background contribution to the ``positive'' effect, the sensitivity of the experiment was simply not enough to detect $^{100}$Mo decay.} and one 
result from a geochemical experiment have been obtained. I do not use the geochemical result here (see comment in Section~\ref{sec.3.3}). Finally, in calculating the average, only the results of experiments with S/B ratios greater than 1 were used (i.e., the results of Refs. \cite{DAS95,DES97,ARN19,CAR14,ARM20}).  I use only final result of Elliott et al. \cite{DES97}  and do not consider their preliminary result from  \cite{ELL91}.   
For $^{100}$Mo SSD mechanism was installed and in Ref.~ \cite{DAS95,ARN19,ARM20} the half-lives were obtained taking this fact into account. In addition, the corrected 
half-life value from Ref. \cite {DAS95} has been used (see explanation in \cite{BAR10}).  
The following weighted average value 
for the half-life is obtained as:
$$
T_{1/2} = 7.06^{+0.15}_{-0.13} \cdot 10^{18} \rm{yr}.                                   
$$

\subsection{$^{100}$Mo - $^{100}$Ru ($0^+_1$; 1130.32 Kev)} 

The $2\nu\beta\beta$ decay of $^{100}$Mo
to the $0^+_1$ excited state of $^{100}$Ru was detected in seven 
independent experiments.  The results are in good agreement. The 
weighted average value for the half-life has been obtained using the results from \cite{BAR95,BAR99,KID09,ARN07,BEL10,ARN14}:
$$
T_{1/2} = 6.7^{+0.5}_{-0.4} \cdot 10^{20} \rm{yr} .
$$     
                              
The result from \cite{KID09}
was used as the final result of the TUNL-ITEP 
 experiment (the result from~\cite{DEB01} was not used here because I consider it as preliminary one). 

\subsection{$^{116}$Cd}
 
Five independent positive 
results were obtained \cite{ARN17,BAR18,EJI95,DAN03,ARN96}. The results are in good agreement with each other.
The corrected 
result for the half-life value from Ref. \cite{ARN96} is used here. The original 
half-life value was decreased by $\sim$25\% (see explanation in \cite{BAR10}). In Refs. \cite{ARN96} and \cite{ARN17}, half-life values were obtained with the assumption that the SSD mechanism was realized. The 
weighted average value is: 
$$          
T_{1/2} = (2.69 \pm 0.09)\cdot 10^{19} \rm{yr}.
$$

\subsection{$^{128}$Te and $^{130}$Te} 

There are a large number of geochemical results for these isotopes. Although the half-life ratio for these isotopes is well known (accuracy is $\sim$  3\% \cite{BER93}), the absolute $T_{1/2}$ values for each isotope are different from one experiment to the next.
One group of authors 
\cite{MAN91,TAK66,TAK96} gives $T_{1/2} \approx 0.8\cdot 10^{21}$ yr  
for $^{130}$Te and $T_{1/2} \approx  2\cdot 10^{24}$ yr for $^{128}$Te, 
while another group \cite{KIR86,BER93} claims $T_{1/2} \approx 
(2.5-2.7)\cdot 10^{21}$ yr and  $T_{1/2} \approx 7.7\cdot 10^{24}$ yr, 
respectively. In addition, as a rule, experiments with young 
samples ($\sim100$ million years) give results of the half-life value for 
$^{130}$Te in the range of $\sim (0.7-0.9)\cdot 10^{21}$ yr,
while experiments with old samples ($> 1$ billion years) give half-life values in the
range of $\sim (2.5-2.7)\cdot 10^{21}$ yr. 
In 2008, it was demonstrated that short half-lives are more likely to be correct \cite{MES08,THO08}.
In a new experiment with young minerals, the half-life values were estimated at 
$(9.0 \pm 1.4)\cdot 10^{20}$ yr \cite{MES08} and $(8.0 \pm 1.1)\cdot 10^{20}$ yr \cite{THO08} 
for $^{130}$Te and $(2.41 \pm 0.39)\cdot 10^{24}$ y \cite{MES08} and $(2.3 \pm 0.3)\cdot 10^{24}$ yr \cite{THO08} 
for $^{128}$Te. In fact, in both experiments, the half-life was measured only for $^{130}$Te, and the value for $^{128}$Te was determined using the previously measured $T_{1/2}(^{130}{\rm Te})/T_{1/2}(^{128}{\rm Te})$ ratio \cite{BER93}. If we average the values obtained in these two experiments, we get:
$T_{1/2} = (8.4 \pm 0.9)\cdot 10^{20}$ years for $^{130}$Te and $T_{1/2} = (2.3 \pm 0.3) \cdot 10^{24}$ years for $^{128}$Te, which is in good agreement with the results of direct (counter) experiments (see below).

The first indication of the observation of the $2\nu\beta\beta$ decay for $^{130}$Te in a direct experiment was obtained in 
\cite{ARN03}. More accurate and reliable values were obtained later in the NEMO-3 experiment \cite{ARN11}. Very precise results were obtained recently in CUORE-0 \cite{ALD17} and CUORE \cite{NUT20} experiments. 
The~results are in good agreement, and the weighted average value is
$$
T_{1/2} = (7.91 \pm 0.21)\cdot 10^{20} yr.
$$ 

Now, using the very well-known ratio $T_{1/2}(^{130}{\rm Te})/T_{1/2}(^{128}{\rm Te}) =
(3.52 \pm 0.11)\cdot 10^{-4}$ \cite{BER93},
one can obtain the half-life value for $^{128}$Te,
$$
T_{1/2} = (2.25 \pm 0.09)\cdot 10^{24} yr.
$$  

 I recommend using these two results as the most correct and reliable half-life values for 
$^{130}$Te and $^{128}$Te. As one can see now, results of direct and geochemical experiments are in good agreement.

\subsection{$^{136}$Xe}
The half-life value for $^{136}Xe$ was measured in two independent experiments, EXO 
 \cite{ACK11,ACK12,ALB14} 
and Kamland-Zen \cite{GAN16,GAN12a,GAN12}.
To obtain the average value of the half-life, the most accurate results of these experiments obtained in \cite{ALB14} and \cite{GAN16} were used (see Table 1). 
The weighted average value is
$$
T_{1/2} = (2.18 \pm 0.05)\cdot 10^{21} yr.
$$

\subsection{$^{150}$Nd}

The positive results were obtained in three 
independent experiments \cite{ART95,DES97,ARN16a}. The most accurate value was obtained in Ref. \cite{ARN16a}. This value is higher than in Ref. \cite{DES97} ($\sim3\sigma$ difference) and lower than in Ref. \cite{ART95} 
($\sim2\sigma$ difference). Using Equations (2) and the three 
above-mentioned  results, one 
obtains $T_{1/2} = (8.4 \pm 0.5)\cdot 10^{18}$ yr.  Taking into account 
that $\chi^2/(N - 1) > 1$ and S = 2.23 (see Equation (3)), we then obtain:
$$
T_{1/2} = (8.4 \pm 1.1)\cdot 10^{18} \rm{yr}.
$$ 

It can be seen that due to the discrepancy between the $T_{1/2}$ values, one has to increase the error in order to somehow agree on the experimental results. On the other hand, it is clear that the result of the NEMO-3 experiment is today the most accurate and reliable. This is confirmed by the fact that in the NEMO-3 experiment, seven different isotopes were investigated simultaneously
. In addition to $^{150}$Nd, $^{48}$Ca, $^{82}$Se, $^{96}$Zr, $^{100}$Mo, $^{116}$Cd and $^{130}$Te were also studied. For all these isotopes, the results are in good agreement with the results of other experiments. It is natural to assume that the result for $^{150}$Nd is correct too. Therefore, I think that it is necessary to use this value as the most accurate at the~moment:
$$
T_{1/2} = 9.34^{+0.67}_{-0.64}\cdot 10^{18} \rm{yr}.
$$ 

\subsection{$^{150}$Nd - $^{150}$Sm ($0^+_1$; 740.4 Kev)}

There are two positive results for $2\nu\beta\beta$ decay of $^{150}$Nd to the $0^+_1$ excited state of $^{150}$Sm 
\cite{BAR09,KID14} (the preliminary result of Ref. \cite{BAR09} was 
published in Ref. \cite{BAR04}). 
These two results are in good agreement. The weighted average value is:

$$          
T_{1/2} = 1.2^{+0.3}_{-0.2}\cdot 10^{20} \rm{yr}.
$$ 

Recently, the result of a new experiment was presented at MEDEX'19  \cite{KAS19} (see Table \ref{Table1}). 
I am not using this new result in my analysis because this is an ongoing experiment and the result is still preliminary and not yet published.

\subsection{$^{238}$U}  
The two-neutrino decay of $^{238}$U was measured in a single experiment using the radiochemical technique \cite{TUR91}:
$$          
T_{1/2} = (2.0 \pm 0.6)\cdot 10^{21} \rm{yr}.
$$ 

It has to be stressed that for $^{238}$U a ``positive'' result was obtained in only the  experiment. Therefore,~it~is necessary to confirm this result in independent experiments (including direct measurements). Until these confirmations are received, one has to be very careful with this value.

\subsection{$^{130}$Ba (ECEC)}

For $^{130}$Ba, positive results were obtained using the geochemical technique only. In this type of measurement, one can not recognize the different modes. It is clear that exactly the ECEC(2$\nu$) process was detected because other modes are strongly suppressed (see estimations in
\cite{DOM05,SIN07,BAR13}). The first time the positive result for $^{130}$Ba was mentioned was in Ref. \cite{BAR96}, where experimental data of Ref.~\cite{SRI76} were analyzed.
In this paper, a positive result was obtained for one sample of barite 
($T_{1/2} = 2.1^{+3.0}_{-0.8} \cdot 10^{21}$~yr), but for a second sample only the limit was set ($T_{1/2} > 4 \cdot 10^{21}$ yr). Later, more accurate 
half-life values, $(2.2 \pm 0.5) \cdot 10^{21}$ yr \cite{MES01} and 
$(0.60 \pm 0.11)\cdot 10^{21}$ yr \cite{PUJ09}, were measured. 
One can see that the results are in 
strong disagreement.
In \cite{MES17}, the data of \cite{PUJ09} were analyzed and it was shown that subtraction of the contribution of cosmogenic $^{130}$Xe removes the contradiction with the result of \cite{MES01}. \linebreak Finally,~I recommend the following value from \cite{MES01}: 
$$          
T_{1/2} = (2.2 \pm 0.5)\cdot 10^{21} \rm{yr}.
$$   
   
To obtain more reliable and precise half-life values, new measurements are needed (including direct~experiments). 

\subsection{$^{78}$Kr (2$ \it {K}$)}

The first indication of the observation of 2$\it {K}$ capture in $^{78}$Kr was announced in 2013 (the effect is $\sim$ 2.5 $\sigma$), $T_{1/2} = [0.92^{+0.55}_{-0.26} (stat) \pm 0.13 (syst)] \cdot 10^{22}$ years \cite{GAV13}. Then, the same data were analyzed more carefully and a new value was published ($\sim$ 4 $\sigma$), which turned out to be twice as much,
$T_{1/2} = [1.9^{+1.3}_{-0.7}(stat) \pm 0.3(syst)] \cdot 10^{22}$ \cite{RAT17}.
The analysis of the data is quite complicated and it is possible that the systematic error is much larger than the indicated 15\%.

There is one more circumstance that makes me cautious about the result given in \cite{RAT17}. As can be seen from Table 3, in the case of $^{78}$Kr, we are dealing with an anomalously large value of nuclear matrix element. This value is significantly larger than in the case of $^{130}$Ba and $^{124}$Xe (1.8 and 5.4~times, respectively) and exceeds all 13 NME values for $2\nu\beta\beta$ decay (from 1.7 to 17.7 times). Here, it is necessary to take into account that, since the rate of the ECEC process is $\sim$ 15-20\% higher than the 2$\it K$ capture, the NME for the ECEC process in $^{78}$Kr is approximately 1.07--1.1 times greater than for the 2$\it K$ capture. This circumstance only strengthens the contradiction.
In principle, such a large NME is possible, but~looks strange. In any case, confirmation of the result \cite{RAT17} in independent measurements is necessary. Until the confirmation, one has to be very careful with this result.

\subsection{$^{124}$Xe (2k)}

To date, only one positive result has been published for 2$\it {K}$ capture in $^{124}$Xe \cite{APR19}:
$T_{1/2} = [1.8 \pm 0.5(stat) \pm 0.13(syst)] \cdot 10^{22}$ yr.                 
The significance of the effect is only 4.4 $\sigma$. It should also be noted that a limit 
$T_{1/2} > 2.1 \cdot 10^{22}$ yr was obtained in \cite{ABE18}, which formally contradicts the result of \cite{APR19}. Taking~into account errors, there is no real contradiction here. However, it is clear that it is necessary to confirm the result of \cite{APR19} in an independent experiment.

\section{NME Values for Two-Neutrino Double Beta Decay}
\label{sec.4}
Obtained average and recommended half-life values are presented in Table \ref{Table.3} (2-nd column). 
Using these values, one can extract the experimental nuclear matrix elements 
through the relation \cite{KOT12}: 

\begin{equation}
T_{1/2}^{-1} = G_{2\nu}\cdot g_A^4\cdot(m_ec^2\cdot M_{2\nu})^2, 
\end{equation}
 
where $T_{1/2}$ is the half-life value in [yr], $G_{2\nu}$ is the phase space factor in [yr$^{-1}$], 
$g_A$ is the dimensionless axial vector coupling constant 
and  $(m_ec^2\cdot M_{2\nu})$  is the dimensionless nuclear matrix element. 
One has to remember that there are indications that in nuclear
medium the $g_A$ value is reduced in comparison with their free nucleon values (see Section \ref{sec.1}).
Expression (5) is valid for $2\nu\beta\beta$ and ECEC(2$\nu$) processes.

Thereby, following Ref. \cite{KOT12}, it is better
to use the so-called effective NME, 
$\mid M^{eff}_{2\nu}\mid = g_A^2\cdot \mid (m_ec^2\cdot M_{2\nu})\mid$. This value has been calculated for all isotopes.    

\begin{table}[H]

\caption{Half-life and effective nuclear matrix element values for $2\nu\beta\beta$ decay (see Section \ref{sec.4}).}
\label{Table.3}
\centering
\scalebox{1.1}[1.1]{
\begin{tabular}{ccccc}
\toprule
\textbf{Isotope} & \boldmath{$T_{1/2}(2\nu)$}\textbf{, yr} &\textbf{$\mid M^{eff}_{2\nu}\mid$}  & \textbf{$\mid M^{eff}_{2\nu}\mid$}   & \textbf{Recommended}  \\ 
& & \textbf{($G_{2\nu}$ from \cite{KOT12})} & \textbf{($G_{2\nu}$ from \cite{MIR15})} & \textbf{Value} \\
\midrule
$2\nu\beta\beta$: & & & &\\
$^{48}$Ca & $5.3^{+1.2}_{-0.8}\cdot10^{19}$ & $0.0348^{+0.0030}_{-0.0034}$ 
& $0.0348^{+0.0030}_{-0.0034}$ & $0.035 \pm 0.003$ \\
$^{76}$Ge & $(1.88 \pm{0.08}) \cdot10^{21}$ & $0.1051^{+0.0023}_{-0.0024}$ 
& $0.1074^{+0.0024}_{-0.0022}$ & $0.106 \pm 0.004$ \\
$^{82}$Se & $0.87^{+0.02}_{-0.01} \cdot10^{20}$ & $0.0849^{+0.0005}_{-0.0010}$ 
& $0.0855^{+0.0005}_{-0.0010}$ & $0.085 \pm 0.001$ \\
$^{96}$Zr & $(2.3 \pm 0.2)\cdot10^{19}$ & $0.0798^{+0.0037}_{-0.0032}$ 
& $0.0804^{+0.0038}_{-0.0033}$ & $0.080 \pm 0.004$ \\
$^{100}$Mo & $7.06^{+0.15}_{-0.13}\cdot10^{18}$ & $0.2071^{+0.0019}_{-0.0022}$
& $0.2096^{+0.0020}_{-0.0022}$ \\
  &  & $0.1852^{+0.0017^{~(a)}}_{-0.0019}$ & & $0.185 \pm 0.002$ \\
$^{100}$Mo- & $6.7^{+0.5}_{-0.4}\cdot10^{20}$ 
& $0.1571^{+0.0048}_{-0.0056}$ &  $0.1619^{+0.0050}_{-0.0058}$  \\
 $^{100}$Ru$(0^{+}_{1})$ &  & $0.1513^{+0.0047^{~(a)}}_{-0.0053}$ & & $0.151 \pm 0.005$ \\ 
$^{116}$Cd & $(2.69 \pm 0.09)\cdot10^{19}$ & $0.1160^{+0.0020}_{-0.0019}$ 
& $0.1176^{+0.0020}_{-0.0019}$\\
 &  & $0.1084^{+0.0024^{~(a)}}_{-0.0019}$ & & $0.108 \pm 0.003$\\
$^{128}$Te & $(2.25 \pm 0.09)\cdot10^{24}$ & $0.0406^{+0.0008}_{-0.0008}$ 
& $0.0454^{+0.0009}_{-0.0009}$ & $0.043 \pm 0.003$ \\
$^{130}$Te & $(7.91 \pm 0.21)\cdot10^{20}$ & $0.0288^{+0.0004}_{-0.0004}$
&  $0.0297^{+0.0004}_{-0.0004}$ & $0.0293 \pm 0.0009$\\
$^{136}$Xe & $(2.18 \pm 0.05)\cdot10^{21}$ & $0.0177^{+0.0002}_{-0.0002}$ 
& $0.0184^{+0.0002}_{-0.0002}$ & $0.0181 \pm 0.0006$ \\
$^{150}$Nd & $(9.34 \pm 0.65)\cdot10^{18}$ & $0.0543^{+0.0020}_{-0.0018}$ 
& $0.0550^{+0.0020}_{-0.0018}$ & $0.055 \pm 0.003$ \\
$^{150}$Nd- & $1.2^{+0.3}_{-0.2}\cdot10^{20}$
& $0.0438^{+0.0042}_{-0.0046}$ & $0.0450^{+0.0043}_{-0.0048}$ & $0.044 \pm 0.005$\\
$^{150}$Sm($0^{+}_{1}$) & & & &\\
$^{238}$U & $(2.0 \pm 0.6)\cdot10^{21}$ & $0.1853^{+0.0361}_{-0.0227}$ 
& $0.0713^{+0.0139}_{-0.0088}$ & $0.13^{+0.09}_{-0.07}$  \\
ECEC(2$\nu$): & & & &\\
$^{78}$Kr$^{~(b)}$ & $1.9^{+1.3}_{-0.8}\cdot10^{22}$ & $0.2882^{+0.0829}_{-0.0706}$ \cite{KOT13} & 
$0.3583^{+0.1126}_{-0.0822}$ & $0.32^{+0.15}_{-0.11}$ \\
$^{124}$Xe$^{~(b)}$ & $(1.8 \pm 0.5)\cdot10^{22}$ & $0.0568^{+0.0101}_{-0.0650}$ \cite{KOT13} & 
$0.0607^{+0.0107}_{-0.0070}$ & $0.059^{+0.013}_{-0.009}$ \\
$^{130}$Ba  & $(2.2 \pm 0.5) \cdot 10^{21}$ & $0.1741^{+0.0239}_{-0.0170}$ \cite{KOT13} & 
$0.1754^{+0.0241}_{-0.0171}$ & $0.175^{+0.024}_{-0.017}$ \\

\bottomrule
\end{tabular}}
\begin{tabular}{c}
\multicolumn{1}{p{\linewidth-1.5cm}}{\footnotesize \justifyorcenter{$^{(a)}$ Obtained using the SSD model.
$^{(b)}$ Value for 2$\it K$ capture. For the ECEC process, the half-life value will be approximately 15--20\% less, and the NME value approximately 7--10\% higher.}}
\end{tabular}

\end{table} 
\unskip
\vspace{12pt}
The obtained results are presented 
in Table \ref{Table.3} (3-rd and 4-th columns). When calculating, I~used the $G_{2\nu}$ values from Refs. \cite{KOT12} and \cite{MIR15}
(see Table \ref{Table.4}). For $^{130}$Ba, $^{78}$Kr and $^{124}$Xe $G_{2\nu}$ values 
for ECEC transition were taken from \cite{KOT13} and \cite{MIR15}.
These calculations are most reliable and correct
at this moment. 
The results of these calculations 
are in reasonable agreement ($\sim$1--7\%) with three exceptions: for $^{128}$Te ($\sim$20\%), $^{78}$Kr ($\sim$30\%) and $^{238}$U (factor $\sim$7 
).
For $^{238}$U, two different values $14.57\cdot 10^{-21} yr^{-1}$~ \cite{KOT12} 
and $98.51\cdot 10^{-21} yr^{-1}$ \cite{MIR15})
were produced. The situation with calculations for $^{238}$U is clearly unsatisfactory and these calculations should be rechecked.
For $^{100}$Mo, $^{100}$Mo-$^{100}$Ru$(0^{+}_{1})$ and $^{116}$Cd, I used  $G_{2\nu}$ calculated 
in Ref. \cite{KOT12} for the SSD
mechanism. The obtained values for $\mid M^{eff}_{2\nu}\mid$ are given in Table~\ref{Table.3} 
and
these are the most correct values for these isotopes.    
So-called recommended values for $\mid M^{eff}_{2\nu}\mid$ are presented in Table \ref{Table.3} (5-th column)
too. These values were obtained as an average of two values, given in columns 3 and 4. 
The recommended value error is chosen to cover all ranges of values from columns 
3 and 4 (taking into account corresponding errors). For $^{100}$Mo, $^{100}$Mo-$^{100}$Ru$(0^{+}_{1})$ 
and $^{116}$Cd, I recommend to use the values obtained with $G_{2\nu}$ for the SSD mechanism.
  
Therefore, for the majority of isotopes an accuracy for $\mid M^{eff}_{2\nu}\mid$ is on the level
$\sim$1--10\%. For $^{78}$Kr, $^{124}$Xe and $^{130}$Ba, it is $\sim$46\%,
$\sim$22\% and $\sim$14\%, respectively.  This is mainly because of the not precise half-life values obtained for these isotopes. 
The most unsatisfactory situation is for $^{238}$U ($\sim$70\%).  
Main~uncertainty in this case is connected with the accuracy of $G_{2\nu}$.

Recently, in Ref. \cite{SIM18}, an improved formalism of the $2\nu\beta\beta$ decay rate was presented, which takes into account the dependence of energy denominators on lepton energies via the Taylor expansion. As~a result, the formula for the half-life starts to be more complicated and contains several different matrix elements and different phase space volumes. That is, a new approach to processing the results will be required.  To do this, some parameters of this approach have to be established from experiment and calculated reliably, e.g., within the interacting shell model (see discussion in \cite{SIM18}).  Nevertheless, the results shown in Table \ref{Table.3} retain their significance since it was demonstrated in \cite{SIM18} that additional terms contribute 
$\sim$3\% to $\sim$25\% to the total decay rate. This means that if we consider expression (5) as the first term of the expansion in the approach \cite{SIM18}, then we can conclude that the values of $\mid M^{eff}_{2\nu}\mid$ obtained in this work give a good estimate for
$g_A^2\cdot$$\mid M^{2\nu}_{GT-1}\mid$ (see Formula (19) in \cite{SIM18}). The~values given in Table \ref{Table.3} overestimate $g_A^2\cdot$$\mid M^{2\nu}_{GT-1}\mid$ values by $\sim$1.5--12\% only, which is comparable to the accuracy of determining $\mid M^{eff}_{2\nu}\mid$. An exception is the situation with the results for $^{100}$Mo and $^{116}$Cd obtained using phase space volumes calculated within the SSD. In this case, the most accurate NME estimate was obtained since the exact value of the energy of the lowest 1$^+$ intermediate state was used in the calculations of the phase space volume.

\begin{table}[H]
\caption{Phase-space factors from Refs. \cite{KOT12}, \cite{MIR15} and \cite{KOT13}. }
\label{Table.4}

\centering
\begin{tabular}{ccc}
\toprule
\textbf{Isotope} & \boldmath{$G_{2\nu} (10^{-21} yr^{-1})$ \cite{KOT12}} & \boldmath{$G_{2\nu} (10^{-21} yr^{-1})$ \cite{MIR15}} \\
\midrule
$2\nu\beta\beta$: & &  \\
\midrule
$^{48}$Ca & 15550 & 15536 \\ 
$^{76}$Ge & 48.17 & 46.47 \\
$^{82}$Se & 1596 & 1573 \\
$^{96}$Zr & 6816 & 6744 \\
$^{100}$Mo & 3308 & 3231 \\
 & $4134^{(a)}$ & \\
$^{100}$Mo-$^{100}$Ru$(0^{+}_{1})$ & 60.55  & 57.08 \\
  & $65.18^{(a)}$ & \\
$^{116}$Cd & 2764 & 2688 \\
  & $3176^{(a)}$ &  \\
$^{128}$Te & 0.2688 & 0.2149 \\
$^{130}$Te & 1529 & 1442 \\
$^{136}$Xe & 1433 & 1332 \\
$^{150}$Nd & 36430 & 35397 \\
$^{150}$Nd-$^{150}$Sm($0^{+}_{1}$) & 4329 & 4116 \\
$^{238}$U & 14.57 & 98.51 \\
\midrule
ECEC(2$\nu$):  &  &   \\
\midrule
$^{78}$Kr   & 0.660  \cite{KOT13} & 0.410  \\
$^{124}$Xe   & 17.200 \cite{KOT13} & 15.096  \\
$^{130}$Ba   & 15.000 \cite{KOT13} & 14.773   \\
\bottomrule
\end{tabular}
\begin{tabular}{c}
\multicolumn{1}{p{\linewidth-1.5cm}}{\footnotesize \justifyorcenter{$^{(a)}${Obtained using SSD model.}}}
\end{tabular}

\end{table}

\section{Conclusions}

Thus, the all positive results for $2\nu\beta\beta$ decay obtained by August 2020 have been analyzed. As~a result, the average values of the half-life were obtained for all considered isotopes. For $^{128}$Te, $^{150}$Nd and $^{130}$Ba, so-called recommended values have also been proposed. Using these obtained average/recommended half-life values, the $\mid M^{eff}_{2\nu}\mid$ values for all considered nuclei were determined.
Finally, previous results from Ref. \cite{BAR15} were successfully updated.
A summary is shown in Table~\ref{Table.3}. 
 I recommend using these values as 
the most correct and reliable currently. If we look at the dynamics of the average values since 2001, we can see that these values were constantly refined over time and did not deviate by more than 1--2 $\sigma$ from the initial value. An exception is the situation with $^{76}$Ge. Here, the average value has steadily increased with time (from $1.42^{+0.09}_{-0.07}\cdot 10^{21}$ yr in 2001 to $(1.88\pm 0.08)\cdot 10^{21}$ yr in 2020). This is due to the low quality of the results obtained in the 1990s. In the latest analysis, the results obtained after 2000 have been used.

At present, $2\nu\beta\beta$  decay was recorded in 11 nuclei, and ECEC capture in 3 nuclei (with some doubts for $^{78}$Kr). The accuracy of determining the half-life for most nuclei lies in the range 2--10\%. It is expected that in the next few years new results will be obtained for $^{76}$Ge (Majorana), $^{100}$Mo~(CUPID-Mo, AMORE, CROSS), $^{116}$Cd (CROSS), $^{130}$Te (CROSS, SNO+), $^{136}$Xe (NEXT-100) and $^{124}$Xe~ (NEXT-100, LUX-ZEPLIN). The final result will be obtained in an experiment to search for the $2\nu\beta\beta$ decay of $^{150}$Nd to the first excited 0$^+$ level of $^{150}$Sm (see \cite{KAS19}). Let us emphasize here the importance of experiments using low-temperature bolometers. In experiments with such detectors, the measurement accuracy of the half-life can reach 1-2\%. At present, such experiments are possible for $^{82}$Se, $^{100}$Mo,~$^{116}$Cd and $^{130}$Te. Apparently, in the future, such measurements will be implemented for $^{48}$Ca as well.  I hope that in the future 2$\beta$ processes will also be found in other nuclei. The search for $2\nu\beta\beta$ processes in $^{124}$Sn, $^{110}$Pd, $^{160}$Gd and the search for ECEC(2$\nu$) processes in $^{96}$Ru, $^{106}$Cd and $^{136}$Ce seem promising. As for the $2\nu\beta\beta$ transitions to the excited states of the daughter nucleus, it seems possible to register a transition to the 0$^+_1$ excited level in measurements with $^{96}$Zr and $^{82}$Se in the near future.

\vspace{6pt} 




\funding{This research was partially funded by Russian Scientific Foundation grant number 18-12-00003.}


\conflictsofinterest{The author declares no conflict of interest.} 





\reftitle{References}


\begin{thebibliography}{999}
\bibitem{GOE35}
Goeppetr-Mayer, M. Double beta-disintegration. {\it Phys. Rev.} {\bf 1935}, {\it 48}, 512--516. 
\bibitem{ARN06}
Arnold, R; Augier, C.; Baker, J.; Barabash, A.S.; Brudanin, V; Caffrey, A.J.; Caurier, E.; Egorov, V.; Errahmane, K.; Etienvre, A.I.; {et al}. Limits on different Majoron decay modes of $^{100}$Mo and $^{82}$Se for neutrinoless double beta decays in the NEMO-3 experiment. {\it Nucl. Phys. A} {\bf 2006}, {\it 765}, 483--494.
\bibitem{ARN19}
Arnold. R.; Augier, C.; Barabash, A.S.; Basharina-Freshville, A.; Blondel, S.; Blot, S.;  Bongrand, M.; Boursette, D.; Brudanin, V.; Busto, J.; {et al}. Detailed studies of $^{100}$Mo two-neutrino double beta decay in
NEMO-3. {\it Eur. Phys. J. C} {\bf 2019}, {\it 79}, 440. 
\bibitem{BAR07}
Barabash, A.S.; Dolgov, A.D.; Dvornicky, R.; Simkovic, F.; Smirnov, A.Yu. Statistics of neutrinos and the double beta decay. {\it Nucl. Phys. B} {\bf 2007}, {\it 783}, 90--111.
\bibitem{ALB16}
Albert, J.B.; Barbeau, P.S.; Beck, D.; Belov, V.; Breidenbach, M.; Brunner, T.; Burenkov, A.; Cao, G.F.; Chambers,  C.;  Cleveland, B.;
 {et al}. First search for Lorentz and CPT violation in double beta decay with EXO-200. {\it Phys. Rev. D} {\bf 2016}, {\it 93}, 072001.
\bibitem{AZO19a}
Azzolini, O.; Beeman, J.W.; Bellini, F.; Beretta, M.; Biassoni, M.; Brofferio, C.; Bucci, C.; Capelli, S.;  Cardani, L.; Carniti, P.; {et al}. First search for Lorentz violation in double beta decay
with scintillating calorimeters. {\it Phys. Rev. D} {\bf 2019}, {\it 100}, 092002.


\bibitem{DEP20}
Deppisch, F.F.; Graf, L; Simkovic, F. Searching for New Physics in two-neutrino double beta decay. \emph{arXiv} {\bf 2020}, arXiv:2003.11836.
\bibitem{DEP20a}
Deppisch, F.F.; Graf, L.; Rodejohann, W; Xu, X.-J. Neutrino self-interactions and double beta decay.\emph{ arXiv} {\bf 2020}, arXiv:2004.11919. 
\bibitem{ING50}
Inghram, M.G.; Reynolds, J.H. Double beta-decay of $^{130}$Te. {\it Phys. Rev.} {\bf 1950}, {\it 78}, 822.
\bibitem{ELL87}
Elliott, S.R.; Hahn, A.A.; Moe, M.K. Direct evidence for two-neutrino double-beta decay in $^{82}$Se. {\it Phys. Rev. Lett.} {\bf 1987}, {\it 59}, 2020--2023.
\bibitem{MES01}
Meshik, A.P.; Hohenberg, C.M.; Pravdivtseva, O.V.; Kapusta, Y.S. Weak decay of
$^{130}$Ba and $^{132}$Ba: Geochemical measurements.
{\it Phys. Rev. C} {\bf 2001}, {\it 64}, 035205.
\bibitem{APR19}
Aprile, E.; Aalbers, J.; Agostini, F.; Alfonsi, M.;  Althueser, L.; Amaro , F.D.; Anthony, M.; Antochi,   V.C.; Arneodo, F.;  Baudis, L.; {et al}. Observation of two-neutrino double electron capture in $^{124}$Xe with XENON1T. {\it Nature} {\bf 2019}, {\it 568}, 532--535.
\bibitem{RAT17}
Ratkevich, S.S.; Gangapshev, A.M.; Gavrilyuk, Yu.M.; Karpeshin, F.F.; Kazalov, V.V.; Kuzminov, V.V.; Panasenko, S.I.; Trzhaskovskaya, M.B.; Yakimenko S.P. Comparative study of the double-{\it K}-shell-vacancy production
in single- and double-electron-capture decay. {\it Phys. Rev. C} {\bf 2017}, {\it 96}, 065502.
\bibitem{BAR10}
Barabash, A.S. Precise half-life values for two-neutrino double-$\beta$ decay. {\it Phys. Rev. C} {\bf 2010}, {\it 81}, 035501.
\bibitem{BAR15}
Barabash, A.S. Average and recommended half-life values
for two-neutrino double beta decay. {\it Nucl. Phys. A} {\bf 2015}, {\it 935}, 52--64.
\bibitem{SIM18}
Simkovoc, F.; Dvornicky, R.; Stefanik, D.; Faessler A. Improved description of the $2\nu\beta\beta$-decay and a possibility to determine the effective axial-vector coupling constant. {\it Phys. Rev. C} {\bf 2018}, {\it 97}, 034315.
\bibitem{ROD06}
Rodin, V.A.; Faessler, A.; Simkovic, F.; Vogel, P. Assessment of uncertainties in QRPA $0\nu\beta\beta$-decay nuclear matrix elements. {\it Nucl. Phys. A} {\bf 2006}, {\it 766}, 107--131.
\bibitem{ROD07}
Rodin, V.A.; Faessler, A.; Simkovic, F.; Vogel, P.  Erratum to: Assessment of uncertainties in QRPA $0\nu\beta\beta$-decay nuclear matrix elements. {\it Nucl. Phys. A} {\bf 2007}, {\it 793}, 213--215.
\bibitem{KOR07}
Kortelainen, M.; Suhonen, J. Nuclear matrix elements of $0\nu\beta\beta$ decay with improved short-range correlations. {\it Phys. Rev. C} {\bf 2007}, {\it 76}, 024315.
\bibitem{SIM08}
Simkovic, F.; Faessler, A.; Rodin, V.; Engel, J. Anatomy of the $0\nu\beta\beta$ nuclear matrix elements. {\it Phys. Rev. C} {\bf 2008}, {\it 77}, 045503.
\bibitem{BER12}
Beringer, J.; Agruin, J.-F.; Barnett, R.M.; Copic, K.; Dahl, O.; Groom, D.E.; Lin, C.-J.; Lys, J.; Murayama, H.; Wohl, C.G.; {et al}. Particle Data Group. {\it Phys. Rev. D} {\bf 2012}, {\it 86}, 010001.
\bibitem{ENG17}
Engel, J.; Menendez, J. Status and future of nuclear matrix elements for neutrinoless double-beta decay: a review. {\it Rep. Prog. Phys.} {\bf 2017}, {\it 80}, 046301.
\bibitem{SUH17}
Suhonen, J. Impact of the quenching of $g_A$ on the sensitivity of $0\nu\beta\beta$ experiments. {\it Phys. Rev. C} {\bf 2017}, {\it 96}, 055501.
\bibitem{SUH19}
Suhonen, J.; Kostensalo J. Double $\beta$ becay and the axial
strength. {\it Frontiers in Physics} {\bf 2019}, {\it 7}, 00029.
\bibitem{KOT12}
Kotila, J.; Iachello, F. Phase-space factors for double-$\beta$ decay. {\it Phys. Rev. C} {\bf 2012}, {\it 85}, 034316.
\bibitem{MIR15}
Mirea, M.; Pahomi, T.; Stoica, S. Values of the phase space factor involved in double beta decay. {\it Rom. Rep. Phys.} {\bf 2015}, {\it 67}, 872--889.
\bibitem{BAR02}
Barabash, A.S. Average (recommended) half-life values for two-neutrino
double-beta decay. {\it Czech. J. Phys.} {\bf 2002}, {\it 52}, 567--573.
\bibitem{BAR06}
Barabash, A.S. Average and recommended half-life values for two-neutrino
double-beta decay: Upgrade'05. {\it Czech. J. Phys.} {\bf 2006}, {\it 56}, 437--445.
\bibitem{BAR19}
Barabash, A.S. Average and recommended half-life values for two-neutrino double beta decay:
Upgrade-2019. {\it AIP Conf. Proc.} {\bf 2019}, {\it 2165}, 020002.
\bibitem{ARN16}
Arnold. R.; Augier, C.; Bakalyarov, A.M.; Baker, J.D.; Barabash, A.S.; Basharina-Freshville, A.; Blondel, S.; Blot, S.;  Bongrand, M.; Brudanin, V.; {et al}. Measurement of the double-beta decay half-life and search
for the neutrinoless double-beta decay of $^{48}$Ca
with the NEMO-3 detector. {\it Phys. Rev. D} {\bf 2016}, {\it 93}, 112008.
\bibitem{AGO15}
Agostini, M.; Allardt, M.; Bakalyarov, A.M.; Balata, M.; Barabanov, I.; Barros, N.; Baudis, L.;   Bauer, C.; Becerici-Schmidt, N.; Bellotti, E.; {et al}. Results on $\beta\beta$ decay with emission of two neutrinos or Majorons in $^{76}$Ge from GERDA Phase I. {\it Eur. Phys. J. C} {\bf 2015}, {\it 75}, 416.
\bibitem{ARN18}
Arnold. R.; Augier, C.; Barabash, A.S.; Basharina-Freshville, A.; Blondel, S.; Blot, S.;  Bongrand, M.;  Boursette, D.; Brudanin, V.; Busto, J.; {et al}. Final results on $^{82}$Se double beta decay to the ground state of $^{82}$Kr from the NEMO-3 experiment. {\it Eur. Phys. J. C} {\bf 2018}, {\it 78}, 821.
\bibitem{AZZ19}
Azzolini, O.; Beeman, J.W.; Bellini, F.; Beretta, M.; Biassoni, M.; Brofferio, C.; Bucci, C.; Capelli, S.;  Cardani, L.; Carniti, P.; {et al}. Evidence of single state dominance in the two-neutrino double-$\beta$ decay of $^{82}$Se with CUPID-0. {\it Phys. Rev. Lett.} {\bf 2019}, {\it 123}, 262501.
\bibitem{ARM20}
Armengaud E.; C. Augier, C.; Barabash, A.S.; Bellini, F.; Benato, G.; Benoît, A.; Beretta, M.; Berge, L.; Billard, J.; Borovlev, Yu.A.;  {et al}. Precise measurement of $2\nu\beta\beta$ decay of $^{100}$Mo with the CUPID-Mo detection technology. { \it Eur. Phys. J. C} {\bf 2020}, {\it 80}, 674.
\bibitem{ARN17}
Arnold. R.; Augier, C.; Baker, J.D.; Barabash, A.S.; Basharina-Freshville, A.; Blondel, S.; Blot, S.; Bongrand, M.; Brudanin, V.; Busto, J.; {et al}. Measurement of the $2\nu\beta\beta$ decay half-life and search for the $0\nu\beta\beta$ decay of $^{116}$Cd with the NEMO-3 detector. {\it Phys. Rev. D} {\bf 2017}, {\it 95}, 012007.
\bibitem{BAR18}
Barabash, A.S.; Belli, P.; Bernabei, R.; Cappella, F.; Caracciolo, V.; Cerulli, R.; Chernyak, D.M.; Danevich, F.A.; d’Angelo, S.; Incicchitti, A.; {et al}. Final results of the Aurora experiment to study 2$\beta$ decay
of $^{116}$Cd with enriched $^{116}$CdWO$_4$ crystal scintillators. {\it Phys. Rev. D} {\bf 2018}, {\it 98}, 092007.
\bibitem{ALD17}
Alduino, C.; Alfonso, K.; Artusa, D.R.; Avignone III, F.T.; Azzolini, O.; Banks, T.I.; Bari, G.; Beeman, J.W.; Bellini, F.; Bersan, A.; {et al}. Measurement of the two-neutrino double-beta decay half-life
of $^{130}$Te with the CUORE-0 experiment. {\it Eur. Phys. J. C} {\bf 2017}, {\it 77}, 13.
\bibitem{NUT20} 
Nutini, I.; Adams, D.Q.; Alduino, C.; Alfonso, K.; Avignone III, F.T.; Azzolini, O.; Bari, G.; Bellini, F.; Benato, G.; Biassoni, M.; {et al}. The Cuore detector and results. {\it J. Low Temp. Phys.} {\bf 2020}, {\it 199}, 519--528.
\bibitem{GAN16}
Gando, A.; Gando, Y.; Hachiya, T.; Hachiya, A.; Hayashida, S.; Ikeda, H.; Inoue, K.; Ishidoshiro, K.; Karino, Y.; Koga, M.; {et al}. Search for Majorana neutrinos near the inverted
mass hierarchy region with KamLAND-Zen. {\it Phys. Rev. Lett.} {\bf 2016}, {\it 117}, 082503.
\bibitem{ARN16a}
Arnold, R.;  Augier, C.; Baker, J.D.; Barabash, A.S.; Basharina-Freshville, A.; Blondel, S.; Blot, S.; Bongrand, M.; Brudanin, V.;  Busto, J.; {et al}. Measurement of the $2\nu\beta\beta$ decay half-life of $^{150}$Nd and a search for $0\nu\beta\beta$ decay
processes with the full exposure from the NEMO-3 detector. {\it Phys. Rev. D} {\bf 2016}, {\it 94}, 072003.
\bibitem{KAS19}
Kasperovych, D.V.; Barabash, A.S.; Belli, P.; Bernabei, R.; Boiko, R.S.; Cappella, F.;  Caracciolo, V.; Cerulli, R.; Danevich, F.T.; Di Marco, A.; {et al}. Study of double-$\beta$ decay of $^{150}$Nd to the first 0$^+$ excited level of $^{150}$Sm. {\it AIP Conf. Proc.} {\bf 2019}, {\it 2165}, 020014.
\bibitem{MES17}
Meshik, A.; Pravdivtseva, O. Weak decay of tellurium and barium isotopes in
geological samples: current status. {\it JPS Conf. Ser.} {\bf 2017}, {\it 14}, 020702.

\bibitem{BAL96}
Balysh, A.; De Silva, A.; Lebedev, V.I.; Lou, K.; Moe, M.K.; Nelson, M.A.; Piepke, A.;  Pronskiy, A.; Vient, M.A.; Vogel, P.; {et al}. Double beta decay of $^{48}$Ca. {\it Phys. Rev. Lett.} {\bf 1996}, {\it 77}, 5186--5189.
\bibitem{BRU00}
Brudanin, V.B.; Rukhadze, N.I.; Briançon, Ch.; Egorov, V.G.; Kovalenko, V.E.; Kovalik, A.; Salamatin, A.V.; Stekl, I.; Tsoupko-Sitnikov, V.V.; Vylov, Ts.; Cermak, P. Search for double beta decay of $^{48}$Ca in the TGV experiment. {\it Phys. Lett. B} {\bf 2000}, {\it 495}, 63--68.
\bibitem{VAS90}
Vasenko, A.A.; Kirpichnikov, V.; Kuznetsov, V.A.; Starostin, A.S.; Djanyan, A.G.; Pogosov, V.S.; Shachysisyan, S.P.; Tamanyan, A.G. New results in the ITEP/YePI double beta decay experiment with enriched germanium detector. {\it Mod. Phys. Lett. A} {\bf 1990}, {\it 5}, 1299--1306.
\bibitem{MIL91}
Miley, H.S.; Avignone, F.T.; Brodzinski, R.L.; Collar, J.I.; Reeves, J.H. Suggestive evidence for the two-neutrino double-$\beta$ decay of $^{76}$Ge. 
{\it Phys. Rev. Lett.} {\bf 1990}, {\it 65}, 3092--3095.
\bibitem{AVI91}
Avignone III, F.T.; Brodzinski, R.L.; Guerard, C.K.; Kirpichnikov, I.V.; Miley, H.S.; Pogosov, V.S.; Reeves, J.H.; Starostin, A.S.; Tamanyan, A.G. Confirmation of the observation of $2\nu\beta\beta$ decay of $^{76}$Ge. {\it Phys. Lett. B} {\bf 1991}, {\it 256}, 559--561.
\bibitem{AVI94}
Avignone, F.T. Double-beta decay: some recent results and developments. {\it Prog. Part. Nucl. Phys.} {\bf 1994}, {\it 32}, 223--245.
\bibitem{MOR99}
Morales, A. Review on double beta decay experiments and comparison with theory. {\it Nucl. Phys. B } {\bf 1999}, {\it 77}, 335--345.
\bibitem{HM03}
Dorr, C; Klapdor-Kleingrothaus, H.V. New Monte-Carlo simulation of the
HEIDELBERG-MOSCOW double beta decay experiment. {\it Nucl. Instr. Meth. A} {\bf 2003}, {\it 513}, 596--621.
\bibitem{ELL92}
Elliott, S.R.; Hahn, A.A.; Moe, M.K.; Nelson, M.A.; Vient, M.A. Double beta decay of $^{82}$Se. {\it Phys. Rev. C} {\bf 1992}, {\it 46}, 1535--1537.
\bibitem{ARN98}
Arnold. R.; Augier, C.; Baker, J.;Barabash, A.S.; Blum, D.; Brudanin, V.; Caffrey, A.J.; Campagne, J.E.; Caurier, E.; Dassie, D.; {et al}. Double-$\beta$ decay of $^{82}$Se. {\it Nucl. Phys. A} {\bf 1998}, {\it 636}, 209--223.
\bibitem{KIR86}
Kirsten, T.; Heusser, E.; Kather, D.; Ohm, J.; Pernicka, E.; Richter, H.  	
\emph{New Geochemical Double Beta Decay Measurements on Various Selenium Ores and Remarks Concerning Tellurium Isotopes}; World Scientific: Singapore, 1986; pp.81--92.
\bibitem{ARN99}
Arnold, R.; Augier, C.; Baker, J.; Barabash, A.; Blum, D.; Brudanin, V.; Caffrey, A.J.; Campagne, J.E.; Caurier, E.; Dassie, D.; {et al}. Double beta decay of $^{96}$Zr. {\it Nucl. Phys. A} {\bf 1999}, {\it 658}, 299--312.
\bibitem{ARG10}
Argyriades, J.; Arnold, R.; Augier , C.; Baker, J.; Barabash, A.S.; Basharina-Freshville, A.; Bongrand, M.; Broudin-Bay, G.; Brudanin, V.; Caffrey, A.J.; {et al}. Measurement of the two neutrino double beta decay
half-life of Zr-96 with the NEMO-3 detector. {\it Nucl. Phys. A} {\bf 2010}, {\it 847}, 168--179.
\bibitem{KAW93}
Kawashima, A.; Takahashi, K.; Masuda, A. Geochemical estimation of the half-life for the double beta decay of $^{96}$Zr. {\it Phys. Rev. C} {\bf 1993}, {\it 47}, R2452-R2456.
\bibitem{WIE01}
Wieser, M.E.; De Laeter, J.R. Evidence of the double $\beta$ decay of zirconium-96 measured in 1.8 $\times$ 10$^9$ year-old zircons. {\it Phys. Rev. C} {\bf 2001}, {\it 64}, 024308.
\bibitem{EJI91}
Ejiri,H.; Fushimi, K.; Kamada, T.; Kinoshita, H.; Yamamoto, N. Double beta decays of Mo-100. {\it Phys. Lett. B} {\bf 1991}, {\it 258}, 17--23.
\bibitem{ELL91}
Elliott, S.R.; Moe, M.K.; Nelson, M.A.; Vient, M.A. The double beta decay spectrum of $^{100}$Mo as measured with a TPC. {\it J. Phys. G} {\bf 1991}, {\it 17}, S145-S153.
\bibitem{DAS95}
Dassie, D.; Eschbach, R.; Hubert, F.; Hubert, Ph.; Isaac, M.C.; Izac, C.; Leccia, F.; Mennrath, P.; Vareille, A.; Longuemare, C.; {et al}. Two-neutrino double-$\beta$ decay measurement of $^{100}$Mo. {\it Phys. Rev. D} {\bf 1995}, {\it 51}, 2090--2100.
\bibitem{ALS97}
Alston-Garnjost, M.; Dougherty, B.L.; Kenney, R.W.; Tripp, R.D.; Krivicich, J.M.; Nicholson, H.W.; Sutton, C.S.; Dieterle, B.D.; Foltz, S.D.; Leavitt, C.P.;
 {et al}. Experimental search for double-$\beta$ decay of
$^{100}$Mo. {\it Phys. Rev. C} {\bf 1997}, {\it 55}, 474--493.
\bibitem{DES97}
De Silva, A.; Moe, M.K.; Nelson, M.A.; Vient, M.A. Double $\beta$ decays of
$^{100}$Mo and $^{150}$Nd. {\it Phys. Rev. C} {\bf 1997}, {\it 56}, 2451--2467.
\bibitem{ASH01}
Ashitkov, V.D.; Barabash, A.S.; Belogurov, S.G.; Carugno, G.; Konovalov, S.I.; Massera, F.; Puglierin, G.; Saakyan, R.R.; Stekhanov, V.N.; Umatov, V.I. Double Beta Decay of $^{100}$Mo. {\it JETP Lett.} {\bf 2001}, {\it 74}, 529--531.
\bibitem{CAR14}
Cardani, L.; Gironi, L.; Ferreiro Iachellini, N.; Pattavina, L.; Beeman, J.W.; Bellini, F.; Casali, N.; Cremonesi, O.; Dafinei, I.; {et al}. First bolometric measurement of the two neutrino double beta decay of $^{100}$Mo with a ZnMoO$_4$ crystals array. {\it J. Phys. G} {\bf 2014}, {\it 41}, 075204.
\bibitem{HID04}
Hidaka, H.; Ly, C.V.; Suzuki, K. Geochemical evidence of the double $\beta$ decay of
$^{100}$Mo. {\it Phys. Rev. C} {\bf 2004}, {\it 70}, 025501.
\bibitem{BAR95}
Barabash, A.S.; Avignone III, F.T.; Collar, J.I.; Guerard, C.K.; Arthur, R.J.; Brodzinski, R.L.; Miley, H.S.; Reeves, J.H.; Meier, J.R.; Ruddick, K.; Umatov, V.I. Two neutrino double-beta decay of $^{100}$Mo to the first excited 0$^+$ state in $^{100}$Ru. {\it Phys. Lett. B} {\bf 1995}, {\it 345}, 408--413.
\bibitem{BAR99}
Barabash, A.S.; Gurriaran, R.; Hubert, F.; Hubert, Ph.; Umatov, V.I. $2\nu\beta\beta$ decay of $^{100}$Mo to the first 0$^+$ excited state in $^{100}$Ru. {\it Phys. At. Nucl.} {\bf 1999}, {\it 62}, 2039--2043.
\bibitem{DEB01}
De Braeckeleer, L.; Hornish, M; Barabash, A.; Umatov, V. Measurement of the $\beta\beta$-decay rate of $^{100}$Mo to the first excited 0$^+$ state of $^{100}$Ru. 
{\it Phys. Rev. Lett.} {\bf 2001}, {\it 86}, 3510--3513.
\bibitem{KID09}
Kidd, M.F.; Esterline, J.H.; Tornow, W.; Barabash, A.S.; Umatov, V.I. New results for double-beta decay of $^{100}$Mo to excited
final states of $^{100}$Ru using the TUNL-ITEP apparatus. {\it Nucl. Phys. A} {\bf 2009}, {\it 821}, 251--261.
\bibitem{ARN07}
Arnold, R.; Augier, C.; Baker, J.; Barabash, A.S.; Bongrand, M.; Broudin, G.; Brudanin, V.; Caffrey, A.J.; Egorov, V.; Etienvre, A.I.; {et al}. Measurement of double beta decay of $^{100}$Mo to excited states in the NEMO 3 experiment. {\it Nucl. Phys. A} {\bf 2007}, {\it 781}, 209--226.
\bibitem{BEL10}
Belli, P.; Bernabei, R.; Boiko, R.S.; Cappella, F.; Cerulli, R.; Danevich, F.A.; d’Angelo, S.; Incicchitti, A.; Kobychev, V.V.; Kropivyansky, B.N.; {et al}. New observation of $2\nu\beta\beta$ decay of $^{100}$Mo to the 0$^+_1$
level of $^{100}$Ru in the ARMONIA experiment. {\it Nucl. Phys. A} {\bf 2010}, {\it 846}, 143--156.
\bibitem{ARN14}
Arnold. R.; Augier, C.; Barabash, A.S.; Basharina-Freshville, A.; Blondel, S.; Blot, S.;  Bongrand, M.; Brudanin, V.; Busto, J.;  Caffrey, A.J.; {et al}. Investigation of double beta decay of $^{100}$Mo to excited
states of $^{100}$Ru. {\it Nucl. Phys. A} {\bf 2014}, {\it 925}, 25--36.
\bibitem{EJI95}
Ejiri, H.; Fushimi, K.; Hazama, R.; Kawasaki, M.; Kouts, V.; Kudomi, N.; Kume, K.; Nagata, K.; Ohsumi, H.; Okada, K.; {et al}. Double beta decays of Cd-116. {\it J. Phys. Soc. Japan} {\bf 1995}, {\it 64}, 339--343.
\bibitem{ARN96}
Arnold. R.; Augier, C.; Barabash, A.S.; Blum, D.; Brudanin, V.;  Campagne, J.E.; Dassie, D.; Egorov, V.;  Eschbach, R.;  Guyonnet, J.l.; {et al}. Double-$\beta$ decay of $^{116}$Cd. {\it Z. Phys. C} {\bf 1996}, {\it 72}, 239--247.
\bibitem{DAN03}
Danevich, F.A.; Georgadze, A.Sh.; Kobychev, V.V.; Kropivyansky, B.N.; Nikolaiko, A.S.; Ponkratenko, O.A.; Tretyak, V.I.; Zdesenko, S.Yu.; Zdesenko, Yu.G. Search for 2$\beta$ decay of cadmium and tungsten isotopes: Final results of the Solotvina experiment. {\it Phys. Rev. C} {\bf 2003}, {\it 68}, 035501.
\bibitem{MAN91}
Manuel, O.K. Geochemical measurements of double-beta decay. {\it J. Phys. G} {\bf 1991}, {\it 17}, S221-S229.


\bibitem{BER93}
Bernatowicz, T.; Brannon, J.; Brazzle, R.; Cowsik, R.; Hohenberg, C.; Podosek, F. Precise determination of relative and absolute $\beta\beta$-decay rates of $^{128}$Te and $^{130}$Te. {\it Phys.~Rev. C} {\bf 1993}, {\it 47}, 806--825.
\bibitem{MES08}
Meshik, A.P.; Hohenberg, C.M.; Pravdivtseva, O.V.; Bernatowicz, T.J.; Kapusta, Y.S. $^{130}$Te and $^{128}$Te double beta decay half-lives. {\it Nucl. Phys. A} {\bf 2008}, {\it 809}, 275--289.
\bibitem{THO08}
Thomas, H.V.; Pattrick, R.A.D.; Crowther, S.A.; Blagburn, D.J.; Gilmour, J.D. Geochemical constraints on the half-life of $^{130}$Te.
{\it Phys. Rev. C} {\bf 2008}, {\it 78}, 054606.
\bibitem{ARN03}
Arnaboldi, C.; Brofferio, C.; Bucci, C.; Capelli, S.;  Cremonesi, O.; Fiorini, E.; Giuliani, A.; Nucciotti, A.;  Pavan, M.; Pedretti M.; {et al}. A calorimetric search on double beta decay of $^{130}$Te. {\it Phys. Lett. B} {\bf 2003}, {\it 557}, 167--175.
\bibitem{ARN11}
Arnold, R.; Augier, C.; Baker, J.; Barabash, A.S.; Basharina-Freshville, A.; Blondel, S.; Bongrand, M.; Broudin-Bay, G.; M.; Brudanin, V.; Caffrey, A.J.; {et al}. A calorimetric search on double beta decay of $^{130}$Te. {\it Phys. Rev. Lett.} {\bf 2011}, {\it 107}, 062504.
\bibitem{ALB14}
Albert, J.B.; Auger, M.; Auty, D.J.; Barbeau, P.S.; Beauchamp, E.; Beck, D.; Belov, V.; Benitez-Medina, C.; Bonatt, J.; Breidenbach, M.; {et al}. Improved measurement of the $2\nu\beta\beta$ half-life of $^{136}$Xe with the EXO-200 detector. {\it Phys. Rev. C} {\bf 2014}, {\it 89}, 015502. 
\bibitem{ART95} 
Artemiev, V.; Brakchman, E.; Karelin, A.; Kirichenko, V.; Klimenko, A.; Kozodaeva, O.; Lubimov, V.; Mitin, A.; Osetrov, S.; Paramokhin, V.; {et al}.  	
Half-life measurement of $^{150}$Nd $2\nu\beta\beta$ decay in the time projection chamber experiment. {\it Phys. Lett. B} {\bf 1995}, {\it 345}, 564--568.
\bibitem{BAR09}
Barabash, A.S.; Hubert, P.; Nachab, A.; Umatov, V.I. Investigation of $\beta\beta$ decay in $^{150}$Nd and $^{148}$Nd to the excited states of daughter nuclei. {\it Phys. Rev. C} {\bf 2009}, {\it 79}, 045501.
\bibitem{KID14}
Kidd, M.F.; Esterline, J.H.; Finch, S.W.; Tornow, W. Two-neutrino double-$\beta$ decay of $^{150}$Nd to excited final states in $^{150}$Sm. {\it Phys. Rev. C} {\bf 2014}, {\it 90}, 055501.
\bibitem{TUR91}
Turkevich, A.L.; T.E. Economou, T.E.; Cowan, G.A. Double beta decay of $^{238}$U. {\it Phys. Rev. Lett.} {\bf 1991}, {\it 67}, 3211--3214.


\bibitem{BAR96}
Barabash, A.S.; Saakyan, R.R. Experimental limits on 2$\beta^+$, $K\beta^+$ and 2{\it K} processes for $^{130}$Ba and on 2{\it K} capture for $^{132}$Ba. {\it Phys. At. Nucl.} {\bf 1996}, {\it 59}, 179--184. 
\bibitem{PUJ09}
Pujol, M.; Marty, B.; Burnard, P.; Philippot, P. Xenon in Archean barite: weak decay of $^{130}$Ba, mass-dependent isotopic fractionation and implication
for barite formation. {\it Geoch. Cosm. Act.} {\bf 2009}, {\it 73}, 6834--6846.
\bibitem{BAR11a}
Barabash A.S.; Brudanin, V.B. Investigation of double-beta decay
with the NEMO-3 detector. {\it Phys. At. Nucl.} {\bf 2011}, {\it 74}, 312--317.
\bibitem{AVI19}
Avignone, F.T.; Elliott, S.R. The search for double beta decay
with germanium detectors: past,
present, and future. {\it Front. Phys.} {\bf 2019}, {\it 7}, 6.
\bibitem{MAN86}
Manuel, O.K. \emph{Geochemical Measurements of Double Beta Decay}; World Scientific: Singapore, 1986; pp.71--80.
\bibitem{DOM05}
Domin, P; Kovalenko, S.; Simkovic, F.; Semenov, S.V. Neutrino accompanied $\beta^{\pm}\beta^{\pm}$, $\beta^+/EC$ and {\it EC/EC}
processes within single state dominance hypothesis. {\it Nucl. Phys. A} {\bf 2005}, {\it 753}, 337--363.
\bibitem{VAS90a}
Vasiliev, S.I,; Klimenko, A.A.; Osetrov, S.B.; Pomanskii, A.A.; Smolnikov, A.A. Observation of the excess of events in the experiment on the search for a two-neutrino double beta decay of $^{100}$Mo. {\it JETP Lett.} {\bf 1990}, {\it 51}, 622--626.
\bibitem{TAK66}
Takaoka N.; Ogata, K. The half-life of Te-130 double beta-decay. {\it Z. Naturforsch} {\bf 1966}, {\it 21}, 84--90.
\bibitem{TAK96}
Takaoka, N.; Motomura, Y.; Nagao, K. Half-life of $^{130}$Te double-$\beta$ decay measured with geologically qualified samples. {\it Phys. Rev. C} {\bf 1996}, {\it 53}, 1557--1561.

\bibitem{ACK11}
Ackerman, N.; Aharmim, B.; Auger, M.; Auty, D.J.; Barbeau, P.S.; Barry, K.; Bartoszek, L.; Beauchamp, E.; Belov, V.; {et al}. Observation of two-neutrino double-beta decay in $^{136}$Xe with the EXO-200 detector. {\it Phys. Rev. Lett.}  {\bf 2011}, {\it 107}, 212501.
\bibitem{ACK12}
Auger, M.; Auty, D.J.; Barbeau, P.S.; Beauchamp, E.; Belov, V.; Benitez-Medina, C.; Breidenbach, M.; Brunner, T.; Burenkov, A.; Cleveland, B.; {et al}. Search for neutrinoless double-beta decay in $^{136}$Xe with EXO-200. {\it Phys. Rev. Lett.}  {\bf 2012}, {\it 109}, 032505.
\bibitem{GAN12}
Gando, A.; Gando, Y.; Hanakago, H.; Ikeda, H.;  Inoue, K.; Kato, R.;  Koga, M.; Matsuda, S.; Mitsui, T.;  Nakada T.; {et al}. Limits on Majoron-emitting double-$\beta$ decays of $^{136}$Xe in the KamLAND-Zen experiment. {\it Phys. Rev. C} {\bf 2012}, {\it 86}, 021601.
\bibitem{GAN12a}
Gando, A.; Gando, Y.; Hanakago, H.; Ikeda, H.;  Inoue, K.; Kato, R.; Koga, M.; Matsuda, S.; Mitsui, T.; Nakada T.; {et al}. Measurement of the double-$\beta$ decay half-life of $^{136}$Xe with the KamLAND-Zen experiment. {\it Phys. Rev. C} {\bf 2012}, {\it 85}, 045504.
\bibitem{BAR04}
Barabash, A.S.; Hubert, F.; Hubert, Ph.; Umatov, V.I. Double beta decay of $^{150}$Nd to the first 0$^+$ excited state of $^{150}$Sm. {\it JETP Lett.} {\bf 2004}, {\it 79}, 10--12.  
\bibitem{SIN07}
Singh, S.; Chandra, R.; Rath, P.K.; Raina, P.K.; Hirsch, J.G. Nuclear deformation and the two-neutrino double-$\beta$ decay in
$^{124,126}$Xe, $^{128,130}$Te, $^{130,132}$Ba and $^{150}$Nd isotopes. {\it Eur. Phys. J. A} {\bf 2007}, {\it 33}, 375--388.
\bibitem{BAR13}
Barea, J.; Kotila, J.; Iachello, F. Neutrinoless double-positron decay and positron-emitting electron capture
in the interacting boson model. {\it Phys. Rev. C} {\bf 2013}, {\it 87}, 057301.
\bibitem{SRI76}
Srinivasan, B. Barites: anomalous xenon from spallation and neutron-induced reactions. {\it Earth Planet. Sci. Lett.} {\bf 1976}, {\it 31}, 129--141.
\bibitem{GAV13}
Gavrilyuk, Yu.M.; Gangapshev, A.M.; Kazalov, V.V.; Kuzminov, V.V.; Panasenko, S.I.; Ratkevich, S.S. Indications of $2\nu$2{\it K} capture in $^{78}$Kr. {\it Phys. Rev. C} {\bf 2013}, {\it 87}, 035501. 


\bibitem{ABE18}
Abe, K.; Hiraide, K.; Ichimura, K.; Kishimoto, Y.; Kobayashi, K.; Kobayashi, M.; Moriyama, S.; Nakahata, M.; Norita, T.;  Ogawa H.; {et al}. Improved search for two-neutrino double electron capture on $^{124}$Xe and $^{126}$Xe using particle
identification in XMASS-I. {\it Prog. Theor. Exp. Phys.} {\bf 2018}, {\it 2018}, 053D03.
\bibitem{KOT13}
Kotila, J.; Iachello, F. Phase space factors for $\beta^+\beta^+$ decay and competing modes of double-$\beta$ decay. {\it Phys. Rev. C} {\bf 2013}, {\it 87}, 024313. 


\end{thebibliography}
\end{document}